\documentclass[conf, onecolumn]{ao4elt3}
\usepackage{graphics}
\usepackage[varg]{txfonts} 
\usepackage[latin1]{inputenc}
\usepackage{multirow}

\begin{document}
\pagestyle{plain} \title{FINAL A\&T STAGES OF THE GEMINI PLANET
  FINDER} \author{Markus Hartung\inst{1}\thanks{mhartung@gemini.edu}
  \and Bruce Macintosh\inst{2}\thanks{macintosh1@llnl.gov} \and Lisa
  Poyneer\inst{2} \and Dmitry Savransky\inst{2} \ \and Don
  Gavel\inst{3} \and Dave Palmer\inst{2} \and Sandrine Thomas\inst{4}
  \and Daren Dillon\inst{3} \and Jeffrey Chilcote\inst{5} \and Patrick
  Ingraham\inst{6} \and Naru Sadakuni\inst{1} \and Kent
  Wallace\inst{7} \and Marshall D. Perin\inst{8} \and Christian
  Marois\inst{9} \and Jerome Maire\inst{10} \and Fredrik
  Rantakyro\inst{1} \and Pascale Hibon\inst{1} \and Les Saddlemyer\inst{9} \and Stephen
  Goodsell\inst{1} }
%
\institute{Gemini Observatory, La Serena, c/o AURA, Casilla 603, Chile
\and Lawrence Livermore National Lab., United States
\and Center for Adaptive Optics, Univ. of California Santa Cruz, United States
\and NASA Ames, Unites States.
\and Univ. of California, Los Angeles, United States
\and Université de Montr\'eal, Canada
\and Jet Propulsion Laboratory, Pasadena, United States
\and Space Telescope Science Institute, Baltimore, United States
\and National Research Council of Canada Herzberg, Victoria, Canada 
\and Dunlap Institute for Astronomy \& Astrophysics, Univ. of Toronto, Canada}
\abstract{ 
The Gemini Planet Imager (GPI) is currently in its final Acceptance \&
Testing stages. GPI is an XAO system based on a tweeter \& woofer
architecture (43 \& 9 actuators respectively across the pupil), with
the tweeter being a Boston Michromachines $64^2$ MEMS device. The XAO
AO system is tightly integrated with a Lyot apodizing
coronagraph. Acceptance testing started in February 2013 at the
University of California, Santa Cruz. A conclusive acceptance review
was held in July 2013 and the instrument was found ready for shipment
to the Gemini South telescope on Cerro Pachon, Chile.  Commissioning
at the telescope will take place by the end of 2013, matching the
summer window of the southern hemisphere. According to current
estimates the 3 year planet finding campaign (890 allocated hours)
might discover, image, and spectroscopically analyze 20 to 40 new
exo-planets.  Final acceptance testing of the integrated instrument
can always bring up surprises when using cold chamber and flexure rig
installations. The latest developments are reported. Also, we will
give an overview of GPI's lab performance, the interplay between
subsystems such as the calibration unit (CAL) with the AO bench.
We report on-going optimizations on the AO controller loop to filter
vibrations and last but not least achieved contrast performance
applying speckle nulling. Furthermore, we will give an outlook of
possible but challenging future upgrades as the implementation of a
predictive controller or exchanging the conventional 48x48 SH WFS with
a pyramid. With the ELT era arising, GPI will proof as a versatile and
path-finding testbed for AO technologies on the next generation of
ground-based telescopes.  }

\maketitle

\section{Introduction}
GPI belongs to the newest generation of XAO direct imaging devices
that will be attached to an 8-10\,m class telescope. Worldwide only a
couple of instruments that are currently built or commissioned have
similar capabilities, i.e. SPHERE (VLT/ESO) \cite{beuzit_spie_2008}
and SCExAO (SUBARU) \cite{jovanovic_ao4elt3_2013}. GPI passed
pre-shipment acceptance successfully in July 2013 and arrived on Cerro
Pachon in Chile in August 2013. The project is schedule-driven and
aims for first light by the end of 2013.
The high contrast capabilities in combination with an integral field
spectrograph (IFS, 2.7''x2.7'' FOV, 192x192 spaxels, 14.3\,mas/spaxel)
\cite{chilcote_spie_2012} allows for the detection of Jupiter mass
planets as close as $\sim0.2$'' (2-3 $\lambda/D$) from the primary as
well as its classification through low resolution spectra (R$\sim40$
in H). Eventually, astronomers can effectively study extra-solar
systems in the 5-40 AU regime where the giant gas planets of our own
solar system reside. This region is almost inaccessible with Doppler
and transit observing techniques.

\section{Contrast}
\begin{figure}[b] \center
\resizebox{0.5\columnwidth}{!}{\includegraphics{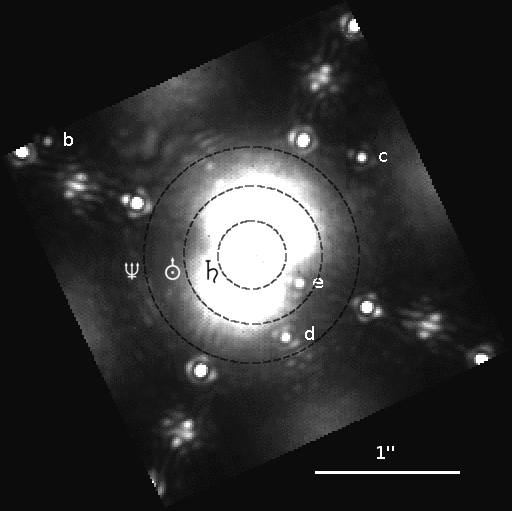}}
\caption{A visual estimate of the appearance of the planets of HR8799
  for 1 min GPI data. The other bright point sources are the reference
  satellite spots created by a diffraction grid imprinted on the
  apodizer masks. Between the first and second order of satellite
  spots a slight waffle pattern is visible. The dashed circles
  correspond to the orbits of Neptun (30\,AU), Uranus (19\,AU) and
  Saturn (9.5\,AU) as seen from 40\,pc, the distance to HR8799.}
\label{fig:hr8799}
\end{figure}

Fig.~\ref{fig:hr8799} shows a slice at 1.6\,$\mu$m of a pipeline
\cite{maire_spie_2012} reconstructed GPI data cube with a 1 min
exposure time. The data correspond to lab measurements on a
calibration source as seen through a telescope simulator optical setup
with a turning phase plate to simulate atmospheric turbulence. As a
demonstration of GPI's expected capabilities the four known planets of
HR8799 were inserted with matching angular dimensions and its PSF peak
fluxes scaled accordingly. The planets can be easily seen on a raw
data cube even without using post processing techniques like
ASDI. Comparable S/N would require about an hour of Keck/NIRC2 time in
broadband imaging, or 5 hours to obtain similar spectral information.

The light of the primary star is suppressed via the well-established
technique of apodized Lyot coronagraphy
\cite{soummer_apj_2005,soummer_spie_2009}. Photometry and reference
PSF are obtained from satellite spots located 0.52''/$\lambda$
off-center (where $\lambda$ is wavelength measured in $\mu$m). These
satellite spots are generated by a grid of fine lines imprinted on the
apodizing masks in a pupil plane following the wavefront correction
surfaces. The ratio between the peak intensity of the
non-coronagraphic PSF (saturating the detector in most cases) to the
peak intensities of the satellite spots is approximately $5 \cdot
10^3$.

\begin{figure}[!ht] \center
\resizebox{0.75\columnwidth}{!}{\includegraphics{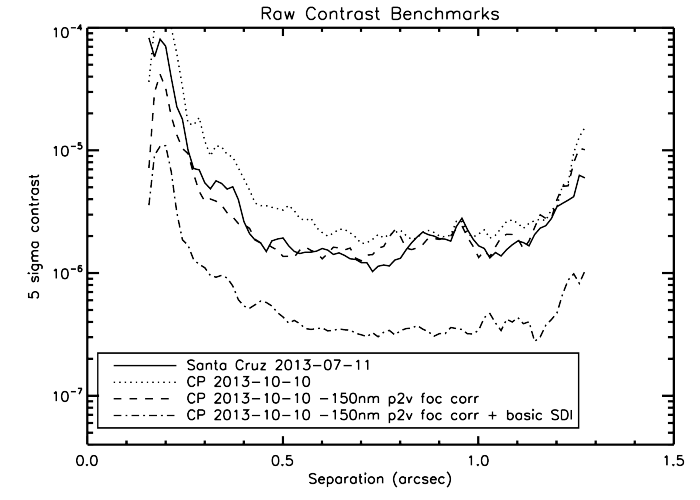} }
\caption{Raw Contrast Benchmarks as measured during pre-shipment
  acceptance in Santa Cruz in July and after delivery on Cerro Pachon
  in October 2013.}
\label{fig:contrast}       
\end{figure}
 
Fig.~\ref{fig:contrast} shows a $5 \sigma$ contrast curve as obtained
at the A\&T location in Santa Cruz compared with first contrast curves
past shipment in the instrument laboratory at Pachon.  Raw contrasts
(no turbulence) as measured after fine-alignment of the instrument
\cite{savransky_ao_2013} approach $10^{-6}$ between 0.5 and
1.0''. After shipment, in the telescope instrument laboratory at Cerro
Pachon, the Santa Cruz benchmark raw contrast values could be
recovered in applying a small focus correction. The figure shows four
contrast curves. The continuous line is the raw contrast as measured
during pre-shipment acceptance. The two other raw contrast curves
correspond to measurements after shipment on Cerro Pachon with and
without a focus correction of 150\,nm P2V, also a good demonstration
of the impact of low order aberrations on final performance (in this
case about a factor of 2 around 0.6''). A basic SDI reduction
(exploiting the fact that speckles move linearly with wavelength but
real objects stay put) brings us well into the $10^{-7}$ regime as
demonstrated by the lowest curve.  We use speckle nulling to tidy up
the floor and to further lower the contrast (not shown
here). Improvements depend on the precision of the used calibrations,
typically up to a factor of two \cite{savransky_spie_2012}, but rather
wavelength depending.

A field of ongoing research and optimization is the control of the
dark hole (for GPI covering the whole FOV for the largest working
wavelength) where the XAO system in combination with its specific
Spatial Filter (SF) settings effectively controls high-order wavefront
aberrations. Our SF is located in the focal plane of an
f/64-beam. Nominally, the SF radius is expected to be $\lambda/(2 d)$
which yields to 2.45\,mm assuming 900\,nm and using 18\,cm
subapertures.  However, it can be difficult to find the best SF size
in broadband light which leads to robust AO performance while passing
through all controllable spatial frequencies.  During A\&T we settled
on a SF size of 2.6\,mm as default. Our supercontinuum light source is
very red, likely one of the reasons why the loop becomes unstable for
a 2.5\,mm setting.

\section{Loops and Alignment}

\begin{table}[b]
\caption{Loops.}
\label{tab:loops}
\begin{center}
\begin{tabular}{ll}
\hline \hline
   & Woofer/Tweeter/TT loop (1\,kHz) \\
   &  Optimal Fourier Controller loop, updates gains \\
   & of the AO controller ($\sim$0.1\,Hz) \\
   & Input Fold Mirror (Gemini pupil - MEMS, $\sim$1\,Hz) \\
\multirow{-5}{*}{\bf{Closed loops (continuous)}} & CAL pointing (via AOWFS P\&C mirrors, $\sim$1\,Hz) \\ \hline
 & CAL/LOWFS to AO WFS ($\sim$0.1\,Hz) \\
 & [HOWFS phase tracking of fringe path length] \\
 & 20 Zernikes to M1/M2 (TLC, filtered 0.1\,Hz, offload 1\,Hz) \\
\multirow{-4}{*}{\bf{Offload loops (continuous)}} & TT/Foc (Gemini synchro bus, filtered 1 Hz, offload 10 Hz) \\ \hline
 & WFS P\&C (MEMS/AOWFS/Apodizer/FPM) \\
\multirow{-2}{*}{\bf{Open loop models (thermal \& flexure)}} & CAL P\&C (IFS Lyot mask/MEMS) \\
\multirow{-2}{*}{-5 to 25 $^\circ$C, -100 to 0 deg zenith angle} & SF translation stage (focal plane) \\
\hline
 & AlignFPM, updates WFS P\&C positions \\
 & (CAL SW/IS, $\sim$1\,Hz) \\ 
\bf{Alignment loops (occasionally)} & Apodizer wheel, centering (IDL gpilib, $\sim$1\,Hz) \\
no human expert judgment required & CAL-IFS P\&C, centering (IDL gpilib, $\sim$1Hz) \\
 & AOWFS to MEMS registration (AOC/IS, $\sim$1Hz) \\
 & Speckle nulling (IDL gpilib, once a minute) \\ \hline
\end{tabular}
\end{center}
\end{table}

Table~\ref{tab:loops} gives an overview of the control loops
\cite{thomas_spie_2012} that are necessary to correct for atmospheric
turbulence, maintain precision alignment \cite{savransky_ao_2013} and
to keep coronagraphic performance at optimum.  At the core of the XAO
system is the fast loop to correct atmospheric turbulence. The spatial
high and low orders are controlled by a tweeter (64x64 Boston
Michromachines MEMS
DM, pupil spans 43 actuators) and a woofer (CILAS 11x11stacked-array
DM, pupil spans 9 actuators), with a straightforward spatial splitting
at the highest woofer Fourier mode $[k, l]$. Valid woofer modes are
selected by the splitting criterion $\sqrt(k^2 + l^2) <= 3.75$ which
results in controlling the woofer up to the modes $[2, 3]$ and $[3,
  2]$. The modes $[3, 3]$ and upwards to the highest mode $[23, 24]$
are on the tweeter. Therefore, the maximum radial frequency on the
woofer is $r = 3.6$ which corresponds to a spatial period of
2.4\,meters when projected onto our telescope pupil.

The woofer is mounted onto a tip/tilt (TT) stage. At 50\,Hz the TT
stage rejection drops by 3\,db, which in practical terms means that
the stage does almost all correction below 20\,Hz and the woofer
surface past 60\,Hz, while the intermediate range is shared by
both. On bright stars GPI is run at 1\,kHz ($\le 8$\,mag) but can be
set to 500\,Hz for fainter stars.

The gain of each Fourier mode is updated by the Optimal Fourier
Controller loop (OFC) \cite{poyneer_oe_2006} every 10 sec. Precise
pupil alignment on the tweeter is ensured by a slow TT control
of a fold mirror at the entrance (input fold mirror), and another
continuous TT loop keeps the star light centered on the Focal Plane
Mask (FPM) inside the calibration unit. This is critical for
coronagraphic performance and has an accuracy requirement of 4\,mas.

\begin{figure}[!ht] \center
\resizebox{0.8\columnwidth}{!}{\includegraphics{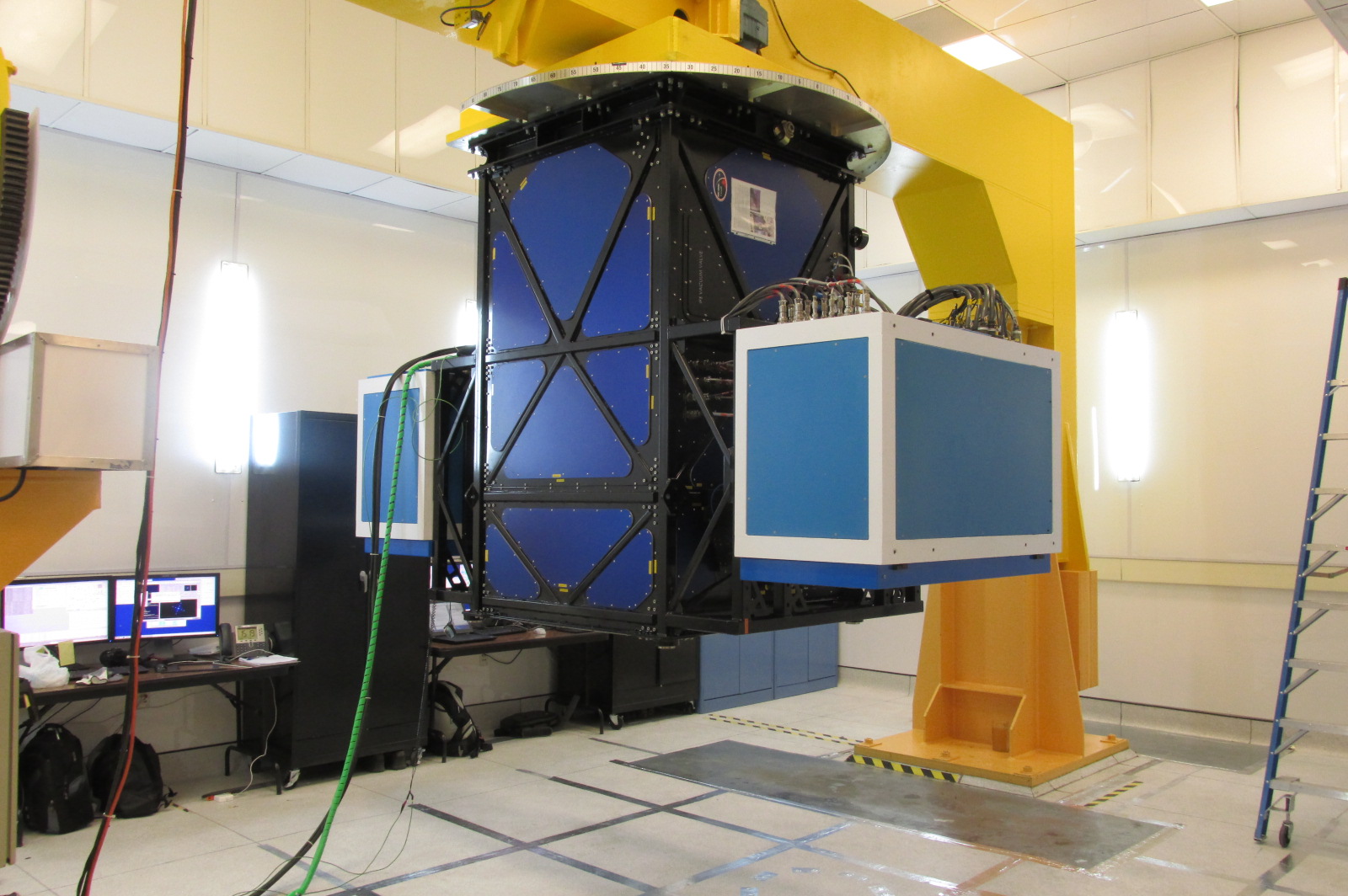}}
\caption{GPI on the flexure mount in the instrument laboratory of the
  Gemini South telescope on Cerro Pachon.}
\label{fig:flexrig}
\end{figure}

The GPI Calibration Unit (CAL) \cite{wallace_spie_2009} is responsible
for the control of non-common path aberrations (NCPA) as measured by
its two WFS. The Low Order WFS (LOWFS) is a 7x7 near-infrared
Shack-Hartmann, the High Order WFS (HOWFS) a Mach-Zehnder type
interferometer. The reconstructed and merged wavefronts are applied as
reference centroid offsets to the main AO loop once a second. For the
time-being the HOWFS functionalities are de-scoped\footnote{Still a
  significant commissioning and development effort is required to
  properly combine the phases. Furthermore, internal vibration reduces
  the contrast of the HOWFS measured fringes more than originally
  expected.} and thus marked with square brackets in the loop overview
(Tab.~\ref{tab:loops}). The first 20 Zernikes will be offloaded to the
telescope mirrors M1 and M2.  Thermal drifts and flexure (GPI is
mounted to a Cassegrain focus) are corrected by open loop models. They
keep the different pupil planes (AOWFS, MEMS, Apodizer, cold IFS Lyot
mask) registered and stabilize GPI contrast performance.  These open
loop models have been established and verified with a dedicated
flexure rig (Fig.~\ref{fig:flexrig}) and cold chamber installations
during the A\&T period. Flexure tests have been repeated after
shipment on Cerro Pachon.

\section{Vibrations}
Mitigating excess vibration is one of the most challenging tasks. The
requirements are tight and contrast performance strongly depends on
this. The residual TT as measured on the wavefront sensor of the
AO module should be less than 4\,mas RMS (for a bright star, excluding
measurement noise).
Obtaining this level of residual TT was challenging in the Santa
Cruz A\&T laboratory environment where the main excitation originated
from the two closed cycle cryocoolers (CryoTel GT from Sunpower Inc.)
attached to the IFS. Once mounted on the telescope, we expect the
vibration environment to change significantly as new sources will be
introduced such as: pumps from other instruments, fans and structural
resonance frequencies excited by the wind, etc.

\begin{figure}[!ht] \center
\resizebox{0.8\columnwidth}{!}{\includegraphics{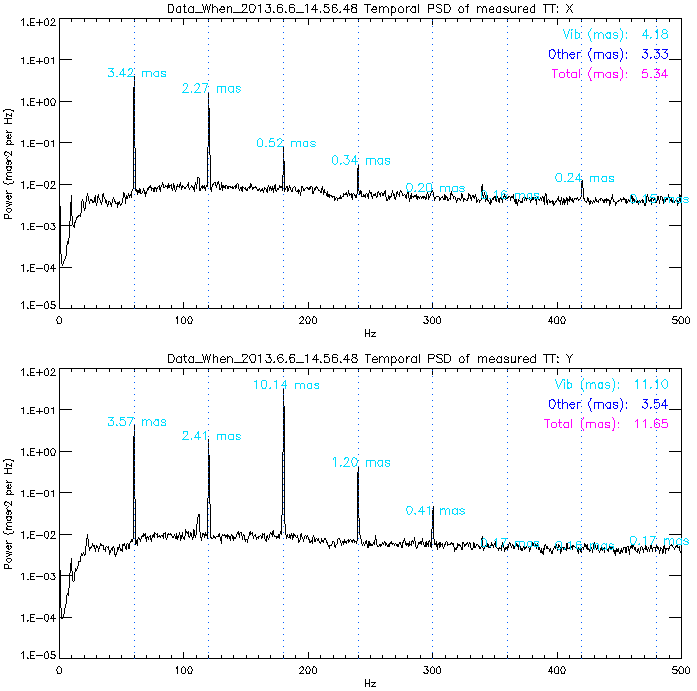} }
\caption{Temporal PSDs of the residual TT as seen by the WFS (mag 7.5,
  SF = 2.8\,mm). In X direction, the total powers (RMS) in the
  vibration frequencies is 4.2\,mas, whereas the remaining 3.3\,mas of
  power comes from other sources. In the Y direction, the power in the
  vibration frequencies is 11.1\,mas, with 3.5\,mas of power coming
  from other sources. The vibration in Y exceeds specification and has
  to be Kalman filtered, particularly for the lines in the overshoot
  region where sensed vibrations are amplified.}
\label{fig:res_tt}
\end{figure}

Fig.~\ref{fig:res_tt} shows typical temporal power spectral densities
for tip and tilt (x, y) using AO telemetry data on a 15\,cm phase
plate\footnote{Atmospheric turbulence was simulated using a rotating
  phase plate \cite{rampy_ao_2012} with $r_0=15$\,cm, wind speed
  2.5\,m/s. If phase plate TT power deviated from Kolmogorov, the TT
  residuals were scaled according RMS[TT phase plate] / RMS[Kolmogorov
    theory].}. For this measurement, the gain of the integral
controller was set to 0.2, all loops were closed (TT, woofer, tweeter)
and the spatial filter was adjusted to a conservative 2.8\,mm. The
Optimal Fourier Controller was running. Note that for bright sources,
the gains of all Fourier modes are limited to 0.3 as a maximum
value. The tilt (y) vibrations are significantly stronger then tip
(x). This reflects the excitation geometry of the two cryocoolers and
the AO WFS. The RMS sum of the vibrations in this figure surpass the
specified 4\,mas RMS, with significant power located at 180\,Hz; well
in the overshoot region of the controller.

To bring the residuals within or close to the specification, a
linear-quadratic-Gaussian (LQG) controller \cite{poyneer_josaa_2010}
is applied to selectively reject vibration lines. These lines may also
be in the overshoot region. The moving components of the Stirling
cryocoolers excite at 60\,Hz but as depicted in Fig.~\ref{fig:res_tt}
its harmonics (120, 180, 240\,Hz) may contain more power. Currently,
we can choose between 27 different vibration filter modes (taking into
account to star magnitude, wind speed, turbulence levels, controllers,
SNR) to optimally adjust for the vibrational conditions. We can
simultaneously correct common-path vibrations and have the ability to
ignore non-common-path vibrations, if required. One challenge is to
achieve a stable loop performance when lines are suppressed
aggressively.  R\&D is still ongoing in this area.

Minor discrepancies between the oscillation frequencies of the two
cryocoolers result into a beating pattern and the vibration strength
varies over a period of approximately 25\,min.  Worse conditions
prevail when the coolers' internal components oscillate in phase. It
should be mentioned Sunpower now provides a hardware mitigation of
this issue via new controller boards that precisely keep the
cryocoolers in opposite phase and prevent any beating. For the time
being, in addition to the LQG filters, we have only implemented a
passive vibration dampening by installing TVAs (Tunable Vibration
Absorbers)
onto the cryocoolers, from Moog CSA. This mitigation was performed
just before shipment and has influenced the vibration spectrum. Its
effects still need to be analyzed in more detail and the system
re-optimized. An additional hardware mitigation step in the near
future is the installation of tunable mass dampers (TMDs) on effective
locations of the GPI structure. Up to 15 fixed mass TMDs will be
installed or fewer with a customized larger TMD mass. No {\it active}
vibration control using external hardware is foreseen.

\section{Upgrades: Predictive Control \& a Pyramid WFS}

\begin{figure}[b] \center
\resizebox{0.75\columnwidth}{!}{\includegraphics{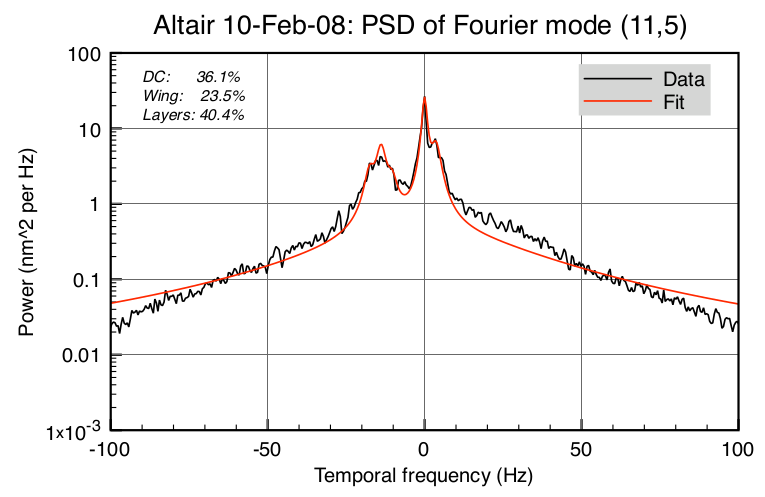} }
\caption{In this example, based on Gemini/Altair telemetry data
  from \cite{poyneer_josaa_2009},  40\% of the power is contained in the the
  frozen flow peaks (layers). }
\label{fig:frozen_flow}       
\end{figure}

The upcoming E-ELT era will challenge current AO technologies. Despite
GPI becoming a facility instrument, there is significant interest to
improve and study the performance of XAO high-contrast systems using
new or recent technologies, such as predictive control or Pyramid
wavefront sensing. From the observatory's standpoint, any impact on
science operations has to be avoided or minimized.  Therefore, an
upgrade of GPI to a predictive control scheme
\cite{poyneer_josaa_2007} is particularly promising since it involves
no extra hardware and exploits the frozen flow hypothesis for
atmospheric turbulence \cite{poyneer_josaa_2009}. This would increase
GPI limiting magnitude by approximately 1 magnitude resulting in a
significant increase the available scientific targets.

Here, we briefly describe the basic recipe
\cite{poyneer_josaa_2007}. The valid actuators in the pupil are mapped
(NxN grid) and the actuator signals are converted to (complex-valued)
modal Fourier coefficients (DFT). The DM influence function is removed
and the actuator signal converted to a phase in nm. The PSD is
calculated for each time series of the complex-valued modal
coefficients. We then employ the property that a pure translation of
Fourier modes produces a peak in temporal frequencies $f_t$ set by the
spatial frequency $f$ and the wind velocity $v$:
\begin{equation}
f_t = v_x f_x + v_y f_y
\end{equation}

\begin{figure}[t] \center
\resizebox{0.6\columnwidth}{!}{\includegraphics{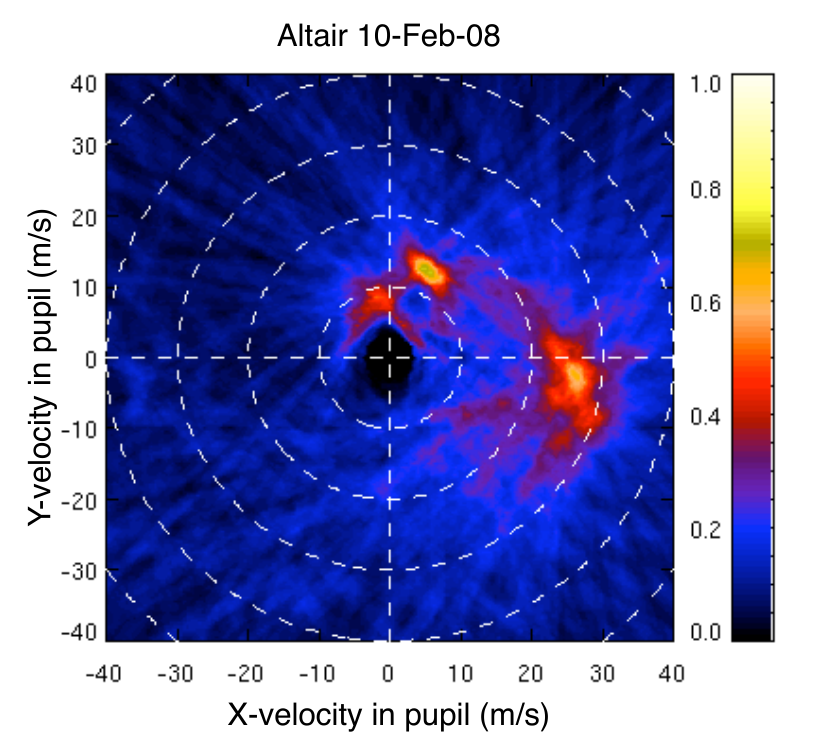} }
\caption{An example of a frozen flow likelihood map. Three layers are
  identified and their locations reveal the corresponding wind
  velocities and directions. Gemini/Altair telemetry data. Figure from
  \cite{poyneer_josaa_2009}.}
\label{fig:likelihood_map}       
\end{figure}

An example of corresponding, ``frozen flow peaks,'' is shown in
Fig.~\ref{fig:frozen_flow}. However, one must be certain that the frozen
flow is indeed detected and this effect is not due to vibration or other unknown phenomena. This
is accomplished by using the frequencies found in {\it all} Fourier modes
to calculate the likelihood for a frozen flow layer using the
percentage of modes belonging to a detected peak (e.g. to +/- 0.75
Hz). This method is very robust at rejecting false positives. For
example, vibrations and aliasing have a likelihood below 5\%. A
corresponding likelihood map is displayed in
Fig.~\ref{fig:likelihood_map}.

Pyramid wavefront sensing has been proven to optimally suit XAO and
high-contrast needs
\cite{guyon_apj_2005,esposito_aa_2013,close_apj_2013,close_apj_2012}.
Upgrading the GPI WFS to a pyramid design was envisioned many years ago, however, no
practical steps have been undertaken to facilitate its implementation. 
Space and weight limitations imposed by GPI being at a Cassegrain focus provide challenging restrictions. Therefore, installing pyramid wavefront sensing
capabilities in the near future is not possible without disrupting science operations. But developers never stop dreaming, and it is
conceivable that after the completion of the GPI planet finding
campaign, the current AO WFS could be replaced by a pyramid during a longer shut
down. If installed, this would improve the limiting magnitude of targets and further increase coronagraphic performance.

\nocite{macintosh_spie_2008}
\nocite{macintosh_spie_2012}

\bibliographystyle{abbrv}

\end{document}